# Asteroseismology: Oscillations on the star Procyon

[Brief Communications Arising]

**François Bouchy**[*], **André Maeder**[†], **Michel Mayor**[†], **Denis Mégevand**[†], **Francesco Pepe**[†], **Danuta Sosnowska**[†]

[*]*Laboratoire d'Astrophysique de Marseille, Traverse du Siphon, BP8, 13372 Marseille cedex 12, France, e-mail : Francois.Bouchy@oamp.fr*

[†]*Observatoire de Genève, 1290 Sauverny, Switzerland*

**Arising from:** J.M. Matthews *et al. Nature* **430**, 51-53 (2004)

Stars are sphere of hot gas whose interiors transmit acoustic waves very efficiently. Geologists learn about the interior structure of Earth by monitoring how seismic waves propagate through it and, in a similar way, the interior of a star can be probed using the periodic motions on the surface that arise from such waves. Matthews *et al.* claim that the star Procyon does not have acoustic surface oscillations of the strength predicted[1]. However, we show here, using ground-based spectroscopy, that Procyon is oscillating, albeit with an amplitude that is only slightly greater than the noise level observed by Matthews et al. using spaced-based photometry.

The new spectrograph HARPS[2] (for High-Accuracy Radial-velocity Planet Searcher), which was installed last year on the 3.6-metre telescope of the European Southern Observatory (La Silla, Chile), was optimised for accuracy in Doppler measurement in order to detect extrasolar planets by means of radial-velocity measurements. During its commissioning, we tested its short-time precision on a sample of bright solar-type stars, including Procyon.

Our measurements on Procyon indicate that there are periodic oscillations of its surface that have a typical period of 18 min (Fig. 1). The apparent amplitude of 4-6 ms$^{-1}$ does not correspond to individual *p*-mode amplitudes, considering that several tens of *p*-modes with similar periods are presumably interfering. Figure 2 presents the Fourier amplitude spectra of the two short time-series obtained on Procyon. No filtering has been applied to the data. Several peaks appear between 0.5 and 1.5 mHz and present the clear signature of acoustic oscillation modes. The correspondence of the main peaks around 1 mHz strongly support the reality of this signature.

Our frequency resolution, which is about 55 µHz, does not allow us to resolve individual *p*-modes. The amplitudes of the peaks between 1 and 1.5 m s$^{-1}$ probably correspond to two or three times the amplitude of individual modes. The mean white-noise level above 2 mHz is respectively 0.11 m s$^{-1}$ and 0.09 m s$^{-1}$ for the first and second sequences. This result, based on only a few hours of observations, confirms and enforces the previous Doppler ground-based detections[3-6].

Why did the Canadian MOST (for Microvariability and Oscillations of Stars) space mission[1] not detect any signatures of *p*-modes on Procyon ? The typical amplitude of *p*-modes is about 0.5 m s$^{-1}$ in radial velocity and the relation for converting between velocity and luminosity amplitudes (given by equation (5) of ref. 7) predicts a luminosity amplitude of only 8-10 p.p.m. This is only slightly greater than the noise level of the satellite obtained after 768 hours of observations. These results indicate that the MOST data are dominated by non-stellar noise, as suspected[8].

However, this conclusion should not overshadow the scientific importance of the Canadian satellite. MOST will lead to breakthroughs on stars with higher oscillation amplitudes, as well as on fast-rotating stars that are not suitable for spectroscopic measurement. The result obtained with HARPS demonstrates the potential of ground-based Doppler measurements for asteroseismology. But for uninterrupted listening to stellar music, a spectrograph like HARPS located in Dome C in Antarctica or in space is needed.

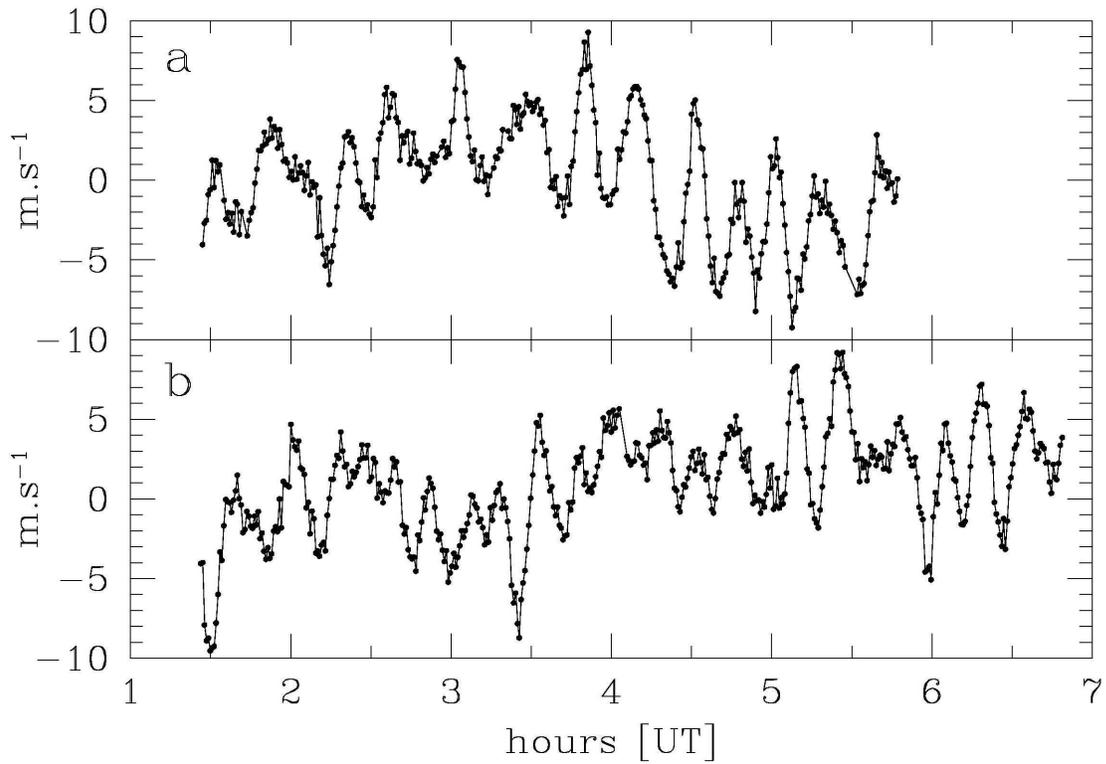

**Figure 1** Short sequences of radial-velocity measurements made on Procyon with HARPS spectrograph. **a, b,** Data were collected on **a,** 5 January 2004 and **b,** 6 January 2004. These sequences indicate oscillation modes with periods of around 18 mn. UT, universal time.

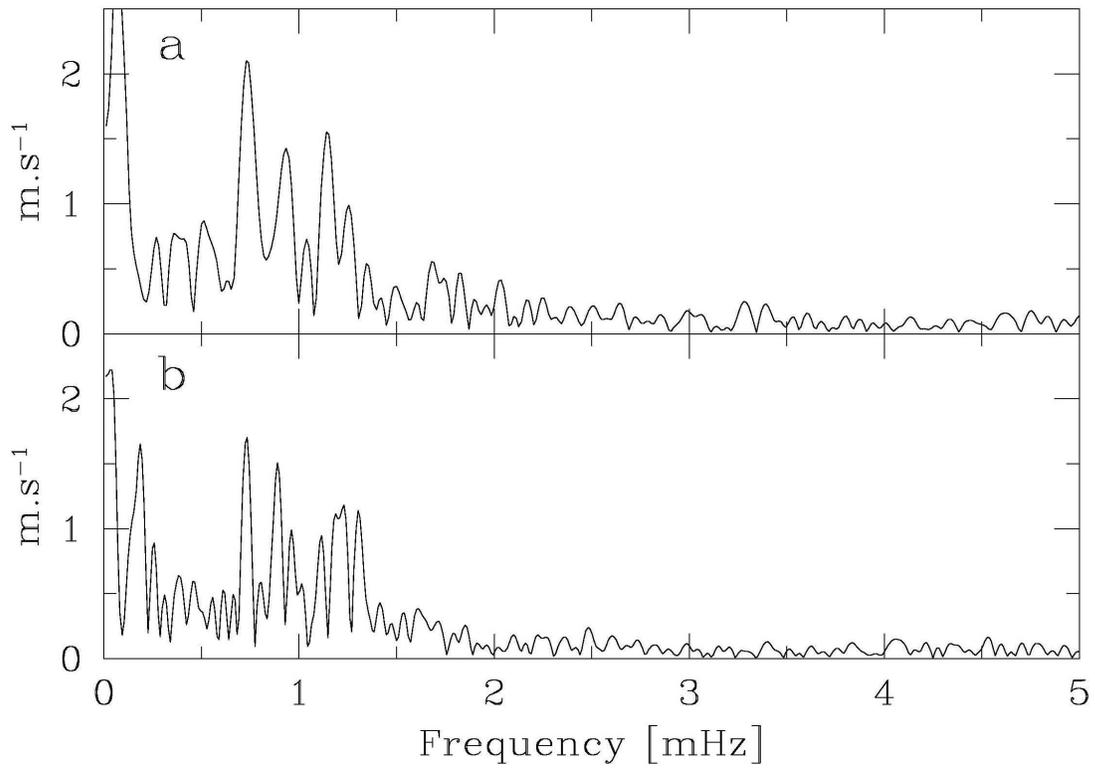

**Figure 2** Fourier amplitude spectra of the two short sequences made on Procyon. **a, b,** Signatures of *p*-modes in the frequency range of 0.5-1.5 mHz are evident.